\documentclass[apjl]{emulateapj}
\usepackage{comment}
\usepackage{ifthen}

\newcommand{\ctbd}[1]{}

\newcommand{\lc}{light curve}
\newcommand{\lcs}{light curves}
\newcommand{\Lc}{Light curve}

\newcommand{\band}[1]{\ensuremath{#1}~band}

\newcommand{\chisq}{\ensuremath{\chi^2}}

\newcommand{\kms}{\ensuremath{\rm km\,s^{-1}}}
\newcommand{\ms}{\ensuremath{\rm m\,s^{-1}}}

\newcommand{\gcmc}{\ensuremath{\rm g\,cm^{-3}}}
\newcommand{\ergscmsq}{\ensuremath{\rm erg\,s^{-1}\,cm^{-2}}}

\newcommand{\teff}{\ensuremath{T_{\rm eff}}}

\newcommand{\vsini}{\ensuremath{v \sin{i}}}
\newcommand{\feh}{\ensuremath{\rm [Fe/H]}}

\newcommand{\vmac}{\ensuremath{v_{\rm mac}}}
\newcommand{\vmic}{\ensuremath{v_{\rm mic}}}
\newcommand{\rhk}{\ensuremath{R^{\prime}_{HK}}}
\newcommand{\logrhk}{\ensuremath{\log\rhk}}

\newcommand{\rsun}{\ensuremath{R_\sun}}
\newcommand{\msun}{\ensuremath{M_\sun}}
\newcommand{\lsun}{\ensuremath{L_\sun}}

\newcommand{\rstar}{\ensuremath{R_\star}}
\newcommand{\mstar}{\ensuremath{M_\star}}
\newcommand{\lstar}{\ensuremath{L_\star}}

\newcommand{\teffstar}{\ensuremath{T_{\rm eff\star}}}
\newcommand{\rhostar}{\ensuremath{\rho_\star}}
\newcommand{\loggstar}{\ensuremath{\log{g_{\star}}}}

\newcommand{\mearth}{\ensuremath{M_\earth}}

\newcommand{\rpl}{\ensuremath{R_{p}}}
\newcommand{\mpl}{\ensuremath{M_{p}}}

\newcommand{\rhopl}{\ensuremath{\rho_{p}}}

\newcommand{\arstar}{\ensuremath{a/\rstar}}
\newcommand{\zrstar}{\ensuremath{\zeta/\rstar}}
\newcommand{\rjup}{\ensuremath{R_{\rm J}}}
\newcommand{\mjup}{\ensuremath{M_{\rm J}}}

\newcommand{\reffig}[1]{Fig.~\ref{fig:#1}}
\newcommand{\refsec}[1]{\mbox{\S\ \ref{sec:#1}}}

\newcommand{\reffigl}[1]{Figure~\ref{fig:#1}}
\newcommand{\refsecl}[1]{\mbox{Section \ref{sec:#1}}}

\newcommand{\reftabl}[1]{Table~\ref{tab:#1}}

\newcommand{\flwof}{\mbox{FLWO 1.2\,m}}

\newcommand{\hatcurhtr}{HTR258-002}
\newcommand{\hatcurfield}{259}
\newcommand{\hatcurCCra}{\ensuremath{03^{\mathrm h}13^{\mathrm m}44.51{\mathrm s}}}
\newcommand{\hatcurCCdec}{\ensuremath{+25{\arcdeg}11{\arcmin}50.7{\arcsec}}}

\newcommand{\hatcurCCtwomass}{2MASS~03134450+2511506}
\newcommand{\hatcurCCgsc}{GSC~1788-01237}
\newcommand{\hatcurCCtassmv}{13.19}
\newcommand{\hatcurCCtwomassJmag}{\ensuremath{11.319\pm0.018}}
\newcommand{\hatcurCCtwomassHmag}{\ensuremath{10.908\pm0.021}}
\newcommand{\hatcurCCtwomassKmag}{\ensuremath{10.815\pm0.018}}

\newcommand{\hatcurCCesoJKmag}{\ensuremath{0.535\pm0.011}}

\newcommand{\hatcurLCdip}{\ensuremath{15.1}}
\newcommand{\hatcurLCrprstar}{\ensuremath{0.1275\pm0.0024}}
\newcommand{\hatcurLCbsq}{\ensuremath{0.208_{-0.073}^{+0.075}}}
\newcommand{\hatcurLCimp}{\ensuremath{0.456_{-0.098}^{+0.073}}}
\newcommand{\hatcurLCzeta}{\ensuremath{19.75\pm0.19}}
\newcommand{\hatcurLCdur}{\ensuremath{0.1174\pm0.0017}}

\newcommand{\hatcurLCdurhr}{\ensuremath{2.817\pm0.041}}

\newcommand{\hatcurLCq}{\ensuremath{0.0321\pm0.0005}}

\newcommand{\hatcurLCingdur}{\ensuremath{0.0163\pm0.0018}}
\newcommand{\hatcurLCP}{\ensuremath{3.652836\pm0.000019}}
\newcommand{\hatcurLCPprec}{\ensuremath{3.6528362}}
\newcommand{\hatcurLCPshort}{\ensuremath{3.6528}}
\newcommand{\hatcurLCT}{\ensuremath{2455176.85173\pm0.00047}}
\newcommand{\hatcurLCTA}{\ensuremath{2454727.55286\pm0.00233}}
\newcommand{\hatcurLCTB}{\ensuremath{2455198.76874\pm0.00050}}
\newcommand{\hatcurLCiblend}{\ensuremath{0.84\pm0.06}}

\newcommand{\hatcurSMEiteff}{\ensuremath{5526\pm88}}
\newcommand{\hatcurSMEizfeh}{\ensuremath{+0.30\pm0.08}}
\newcommand{\hatcurSMEizfehshort}{\ensuremath{0.30}}
\newcommand{\hatcurSMEilogg}{\ensuremath{4.55\pm0.09}}
\newcommand{\hatcurSMEivsin}{\ensuremath{1.4\pm0.7}}
\newcommand{\hatcurSMEivmac}{\ensuremath{3.64}}
\newcommand{\hatcurSMEivmic}{\ensuremath{0.85}}
\newcommand{\hatcurSMEiiteff}{\ensuremath{5500\pm80}}
\newcommand{\hatcurSMEiizfeh}{\ensuremath{+0.31\pm0.08}}
\newcommand{\hatcurSMEiizfehshort}{\ensuremath{+0.31}}
\newcommand{\hatcurSMEiilogg}{\ensuremath{4.46\pm0.06}}
\newcommand{\hatcurSMEiivsin}{\ensuremath{0.5\pm0.5}}
\newcommand{\hatcurSMEiivmac}{\ensuremath{3.60}}
\newcommand{\hatcurSMEiivmic}{\ensuremath{0.85}}

\newcommand{\hatcurDSgamma}{\ensuremath{-12.51\pm0.13}}

\newcommand{\hatcurLBii}{\ensuremath{0.3287}}
\newcommand{\hatcurLBiii}{\ensuremath{0.3039}}

\newcommand{\hatcurISOm}{\ensuremath{1.01\pm0.03}}

\newcommand{\hatcurISOmlong}{\ensuremath{1.010\pm0.032}}
\newcommand{\hatcurISOr}{\ensuremath{0.96_{-0.04}^{+0.05}}}

\newcommand{\hatcurISOrlong}{\ensuremath{0.959_{-0.037}^{+0.054}}}

\newcommand{\hatcurISOlogg}{\ensuremath{4.48\pm0.04}}
\newcommand{\hatcurISOlum}{\ensuremath{0.75\pm0.10}}

\newcommand{\hatcurISOmv}{\ensuremath{5.19\pm0.16}}

\newcommand{\hatcurISOage}{\ensuremath{3.2\pm2.3}}

\newcommand{\hatcurISOMK}{\ensuremath{3.44\pm0.11}}
\newcommand{\hatcurISOJK}{\ensuremath{0.46\pm0.02}}
\newcommand{\hatcurISOspec}{G5}
\newcommand{\hatcurRVK}{\ensuremath{74.3\pm2.4}}
\newcommand{\hatcurRVk}{\ensuremath{0.008\pm0.012}}
\newcommand{\hatcurRVh}{\ensuremath{-0.020\pm0.034}}
\newcommand{\hatcurRVgamma}{\ensuremath{13.5\pm1.8}}
\newcommand{\hatcurRVjitter}{\ensuremath{3.5}}
\newcommand{\hatcurRVeccen}{\ensuremath{0.032\pm0.022}}
\newcommand{\hatcurRVomega}{\ensuremath{271\pm117}}
\newcommand{\hatcurPPi}{\ensuremath{87.6\pm0.5}}

\newcommand{\hatcurPPlogg}{\ensuremath{3.00_{-0.06}^{+0.04}}}
\newcommand{\hatcurPPar}{\ensuremath{10.46_{-0.55}^{+0.38}}}
\newcommand{\hatcurPParel}{\ensuremath{0.0466\pm0.0005}}
\newcommand{\hatcurPPrho}{\ensuremath{0.42\pm0.07}}

\newcommand{\hatcurPPmlong}{\ensuremath{0.567\pm0.022}}

\newcommand{\hatcurPPrlong}{\ensuremath{1.190_{-0.056}^{+0.081}}}

\newcommand{\hatcurPPmrcorr}{\ensuremath{0.15}}
\newcommand{\hatcurPPteff}{\ensuremath{1202\pm36}}
\newcommand{\hatcurPPtheta}{\ensuremath{0.044\pm0.003}}

\newcommand{\hatcurPPfluxavg}{\ensuremath{4.72\pm0.58}}
\newcommand{\hatcurPPfluxavgdim}{\ensuremath{8}}

\newcommand{\hatcurXsecondary}{\ensuremath{2455178.698\pm0.027}}
\newcommand{\hatcurXsecdur}{\ensuremath{0.1138\pm0.0060}}
\newcommand{\hatcurXsecingdur}{\ensuremath{0.0154\pm0.0018}}

\newcommand{\hatcurXdist}{\ensuremath{297_{-13}^{+17}}}
\newcommand{\hatcurEBVtotal}{\ensuremath{0.159\pm0.012}}
\newcommand{\hatcurEBV}{\ensuremath{0.131\pm0.011}}
\newcommand{\hatcurCCesoJKmagf}{\ensuremath{0.465\pm0.014}}

\newcommand{\hatcurlogrhk}{\ensuremath{-4.99}}
\newcommand{\hatcurshk}{\ensuremath{0.166}}
\newcommand{\hatcur}{HAT-P-25}
\newcommand{\hatcurb}{HAT-P-25b}

\newcommand{\hatcurCCtassvi}{\ensuremath{0.53\pm0.12}}
\newcommand{\hatcurSMEversion}{ii}
\newcommand{\hatcurSMEteff}{\ifthenelse{\equal{\hatcurSMEversion}{i}}{\hatcurSMEiteff}{\hatcurSMEiiteff}}
\newcommand{\hatcurSMEzfeh}{\ifthenelse{\equal{\hatcurSMEversion}{i}}{\hatcurSMEizfeh}{\hatcurSMEiizfeh}}
\newcommand{\hatcurSMEzfehshort}{\ifthenelse{\equal{\hatcurSMEversion}{i}}{\hatcurSMEizfehshort}{\hatcurSMEiizfehshort}}
\newcommand{\hatcurSMElogg}{\ifthenelse{\equal{\hatcurSMEversion}{i}}{\hatcurSMEilogg}{\hatcurSMEiilogg}}
\newcommand{\hatcurSMEvsin}{\ifthenelse{\equal{\hatcurSMEversion}{i}}{\hatcurSMEivsin}{\hatcurSMEiivsin}}
\newcommand{\hatcurSMEvmac}{\ifthenelse{\equal{\hatcurSMEversion}{i}}{\hatcurSMEivmac}{\hatcurSMEiivmac}}
\newcommand{\hatcurSMEvmic}{\ifthenelse{\equal{\hatcurSMEversion}{i}}{\hatcurSMEivmic}{\hatcurSMEiivmic}}
\newcommand{\hatcurisoshort}{YY}
\newcommand{\hatcurisofull}{Yonsei-Yale (YY)}
\newcommand{\hatcurisocite}{yi:2001}
\newcommand{\hatcurlumind}{\arstar}
\newcommand{\hatcurjhkfilset}{ESO}

\newboolean{emulateapj}
\setboolean{emulateapj}{true}

\newboolean{astroph}
\setboolean{astroph}{true}

\shortauthors{Quinn et al.}
\shorttitle{\hatcur\lowercase{b}}
\ifthenelse{\boolean{emulateapj}}{
    \newcommand{\titledag}{$\dagger$}
}{
    \newcommand{\titledag}{\dagger}
}

\begin{document}

\title{\hatcur\lowercase{b}: a Hot-Jupiter Transiting a Moderately Faint 
	G Star\altaffilmark{\titledag}}

\author{
	S.~N.~Quinn\altaffilmark{1},
	G.~\'A.~Bakos\altaffilmark{1,2},
	J.~Hartman\altaffilmark{1},
	G.~Torres\altaffilmark{1},
	G.~Kov\'acs\altaffilmark{3},
	D.~W.~Latham\altaffilmark{1},
	R.~W.~Noyes\altaffilmark{1},
	D.~A.~Fischer\altaffilmark{4},
	J.~A.~Johnson\altaffilmark{5},
	G.~W.~Marcy\altaffilmark{6},
	A.~W.~Howard\altaffilmark{6},
	A.~Szentgyorgyi\altaffilmark{1},
	G.~F\H{u}r\'esz\altaffilmark{1},
	L.~A.~Buchhave\altaffilmark{1,7},
	B.~B\'eky\altaffilmark{1},
	D.~D.~Sasselov\altaffilmark{1},
	R.~P.~Stefanik\altaffilmark{1},
	G.~Perumpilly\altaffilmark{1,8},
	M.~Everett\altaffilmark{1},
	J.~L\'az\'ar\altaffilmark{9},
	I.~Papp\altaffilmark{9},
	P.~S\'ari\altaffilmark{9}
}
\altaffiltext{1}{Harvard-Smithsonian Center for Astrophysics,
    Cambridge, MA; email: gbakos@cfa.harvard.edu}

\altaffiltext{2}{NSF Fellow}

\altaffiltext{3}{Konkoly Observatory, Budapest, Hungary}

\altaffiltext{4}{Astronomy Department, Yale University, New Haven, CT}

\altaffiltext{5}{California Institute of Technology, Department of
    Astrophysics, MC 249-17, Pasadena, CA} 

\altaffiltext{6}{Department of Astronomy, University of California,
    Berkeley, CA} 

\altaffiltext{7}{Niels Bohr Institute, Copenhagen University, DK-2100
    Copenhagen, Denmark} 

\altaffiltext{8}{Department of Physics, University of South Dakota,
    Vermillion, SD} 

\altaffiltext{9}{Hungarian Astronomical Association, Budapest,
    Hungary}

\altaffiltext{$\dagger$}{
	Based in part on observations obtained at
	the W.~M.~Keck Observatory, which is operated by the
	University of California and the California Institute of
	Technology. Keck time has been granted by NOAO (A201Hr), NASA
	(N018Hr and N167Hr), and the NASA Gemini-Keck time-exchange
	program (G329Hr).  Based in part on observations made with the
	Nordic Optical Telescope, operated on the island of La Palma
	jointly by Denmark, Finland, Iceland, Norway, and Sweden, in
	the Spanish Observatorio del Roque de los Muchachos of the
	Instituto de Astrofisica de Canarias. 
}

\begin{abstract}
\setcounter{footnote}{9} 
We report the discovery of \hatcurb{}, a transiting extrasolar planet
orbiting the V=\hatcurCCtassmv\ \hatcurISOspec\ dwarf star
\hatcurCCgsc, with a period $P=\hatcurLCP\,d$, transit epoch $T_c =
\hatcurLCT$ (BJD\footnote{Barycentric Julian dates throughout the
paper are calculated from Coordinated Universal Time (UTC)}), and
transit duration \hatcurLCdur\,d.  The host star has a mass of
\hatcurISOm\,\msun, radius of \hatcurISOr\,\rsun, effective
temperature \hatcurSMEteff\,K, and metallicity $\feh =
\hatcurSMEzfeh$.  The planetary companion has a mass of
\hatcurPPmlong\,\mjup, and radius of \hatcurPPrlong\,\rjup\ yielding a
mean density of \hatcurPPrho\,\gcmc.  Comparing these observations
with recent theoretical models, we find that \hatcurb{} is consistent
with a hydrogen-helium dominated gas giant planet with negligible core
mass and age $3.2 \pm 2.3$\,Gyr.  The properties of \hatcurb{} support
several previously observed correlations for planets in the mass range
$0.4<M<0.7\,\mjup$, including those of core mass vs.~metallicity,
planet radius vs.~equilibrium temperature, and orbital period
vs.~planet mass.  We also note that \hatcurb{} orbits the faintest
star found by HATNet to have a transiting planet to date, and is one
of only a very few number of planets discovered from the ground
orbiting a star fainter than V=$13.0$.
\end{abstract}



\section{Introduction}
\label{sec:introduction}

As more transiting extrasolar planets (TEPs) are discovered,
statistics become significant enough to begin looking at bulk
properties of exoplanet populations.  By investigating relationships
between stellar, planetary, and orbital characteristics \citep[see,
e.g.;][]{hartman:2010,burrows:2007,enoch:2010}, we can probe the
underlying astrophysics that dictate the properties we observe.  In
this paper, we present the discovery of \hatcurb{}, the 25th TEP found
by the Hungarian-made Automated Telescope Network
\citep[HATNet;][]{bakos:2004} survey, orbiting the star also known as
\hatcurCCgsc{}.  \hatcurb{} bolsters the population of gas giants in
the range $0.4 < M < 0.7\,\mjup$, and supports several suggestive
correlations between planetary and stellar parameters.

As we observe more stars in search of TEPs, we also expand the
parameter space covered by our discoveries.  One way in which we do so
is by following up candidate planets around fainter stars.  Wide-field
ground-based transit surveys like HATNet have extensive photometric
datasets of faint stars from many different fields, but have tended to
focus their efforts on the brightest stars from each field, as these
are easier to follow up.  Space-based surveys like Kepler forego
picking bright stars from many different fields in order to take
advantage of continuous monitoring of a single field, and thus end up
finding a large percentage of their candidates around faint stars.
Faint stars pose new challenges, both for photometric detection, and
for follow-up to rule out false positives and confirm planetary
status.  With the rich datasets already in hand from ground-based
surveys and the rapidly growing number of faint stars observed from
space, discovering the nuances of faint star follow-up will become
increasingly important.  To date, \hatcur{} is the faintest star
around which HATNet has found a planet, and, while not yet close to
being prohibitively faint, it provides a glimpse at some of the
obstacles that may become routine in the future.

HATNet has been one of the main contributors to the discovery of TEPs.
In operation since 2003, it has now covered approximately 14\% of the
sky, searching for TEPs around bright stars ($8\lesssim I \lesssim
14.0$).  HATNet operates six wide-field instruments: four at the Fred
Lawrence Whipple Observatory (FLWO) in Arizona, and two on the roof of
the hangar housing the Smithsonian Astrophysical Observatory's
Submillimeter Array, in Hawaii.

The layout of the paper is as follows. In \refsecl{obs} we report the
detection of the photometric signal and the follow-up spectroscopic
and photometric observations of \hatcur{}.  In \refsecl{analysis} we
describe the analysis of the data, beginning with the determination of
the stellar parameters, continuing with a discussion of the methods
used to rule out non-planetary, false positive scenarios which could
mimic the photometric and spectroscopic observations, and finishing
with a description of our global modeling of the photometry and radial
velocities.  Our findings are discussed in \refsecl{discussion}.


\section{Observations}
\label{sec:obs}

\subsection{Photometric detection}
\label{sec:detection}

The transits of \hatcurb{} were detected with the HAT-10 telescope in
Arizona and the HAT-9 telescope in Hawaii.  The region around
\hatcurCCgsc{}, a field internally labeled as \hatcurfield, was
observed on a nightly basis between 2008 September 15 and 2009 March
16, whenever weather conditions permitted.  We gathered 6786 exposures
of 5 minutes at a 5.5 minute cadence.  Each image contained
approximately 36,000 stars down to Sloan $r\sim14.5$.  For the
brightest stars in the field, we achieved a per-image photometric
precision of 4\,mmag.

\begin{figure}[!ht]
\plotone{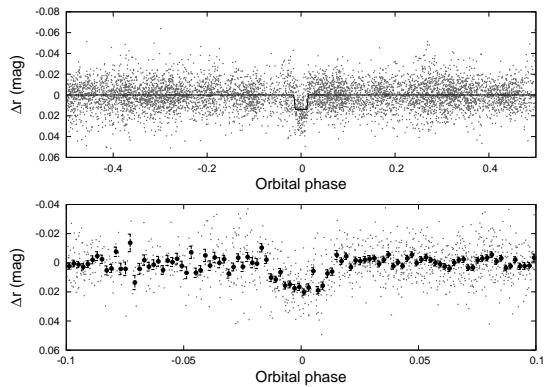}
\caption{
	Unbinned (top) and binned (bottom) \lcs\ of \hatcur{}
        including all 6,786 instrumental Sloan \band{r} 5.5 minute
        cadence measurements obtained with the HAT-9 and HAT-10
        telescopes of HATNet (see the text for details), and folded
        with the period $P = \hatcurLCPprec$\,days resulting from the
        global fit described in \refsecl{analysis}).  The solid line
        shows the ``P1P3'' transit model fit to the light curve
        (\refsecl{globmod}).
\label{fig:hatnet}}
\end{figure}

The calibration of the HATNet frames was carried out using standard
photometric procedures.  The calibrated images were then subjected to
star detection and astrometry, as described in \cite{pal:2006}.
Aperture photometry was performed on each image at the stellar
centroids derived from the Two Micron All Sky Survey
\citep[2MASS;][]{skrutskie:2006} catalog and the individual
astrometric solutions.  The resulting \lcs\ were decorrelated (cleaned
of trends) using the External Parameter Decorrelation \citep[EPD;
see][]{bakos:2009} technique in ``constant'' mode and the Trend
Filtering Algorithm \citep[TFA; see][]{kovacs:2005}.  The \lcs{} were
searched for periodic box-shaped signals using the Box Least-Squares
\citep[BLS; see][]{kovacs:2002} method.  We detected a significant
signal in the \lc{} of \hatcurCCgsc{} (also known as
\hatcurCCtwomass{}; $\alpha = \hatcurCCra$, $\delta = \hatcurCCdec$;
J2000; V=\hatcurCCtassmv{}, \citealp{droege:2006}), with an apparent
depth of $\sim\hatcurLCdip$\,mmag, and a period of
$P=\hatcurLCPshort$\,days (see \reffigl{hatnet}).  The drop in
brightness had a first-to-last-contact duration, relative to the total
period, of $q = \hatcurLCq$, corresponding to a total duration of $Pq
= \hatcurLCdurhr$~hr (see \reffigl{hatnet}).

\subsection{Reconnaissance Spectroscopy}
\label{sec:recspec}

As is routine in the HATNet project, all candidates are subjected to
careful scrutiny before investing valuable time on large telescopes.
This includes spectroscopic observations at relatively modest
facilities to establish whether the transit-like feature in the light
curve of a candidate might be due to astrophysical phenomena other
than a planet transiting a star.  Many of these false positives are
associated with large radial-velocity variations in the star (tens of
\kms) that are easily recognized.  We used two facilities to carry out
the reconnaissance spectroscopy; the Tillinghast Reflector Echelle
Spectrograph \citep[TRES;][]{furesz:2008} on the 1.5\,m Tillinghast
Reflector at FLWO, and the FIbre-fed Echelle Spectrograph
\citep[FIES;][]{frandsen:1999} on the 2.5\,m Nordic Optical Telescope
\citep[NOT;][]{djupvik:2010} at La Palma, Spain.  Both of these
instruments provide high-resolution spectra which, with even modest
signal-to-noise (S/N) ratios, are suitable for deriving RVs with
moderate precision ($\lesssim 0.3\,\kms$) for slowly rotating stars.
We also use the reconnaissance spectra to estimate effective
temperatures, surface gravities, and projected rotational velocities
of the target star.  One or two well timed spectra can very
effectively rule out many types of false positives, including F-M
binaries, grazing eclipsing binaries, or systems of three or more
stars, all of which can be expected to show velocity variation and/or
composite spectra at levels detectable by these observations.
Additional false positive tests are performed with other spectroscopic
material described in the next section.  The reconnaissance
observations are summarized in \reftabl{recon}.  Below we provide a
brief description of each instrument, the data reduction, and the
analysis.

Our first spectrum was taken with the medium fiber on TRES, which has
a resolving power of $\lambda/\Delta\lambda \approx 44,\!000$ and a
wavelength coverage of $\sim$3900--8900\,\AA\@.  The second spectrum
was taken with the medium fiber on FIES, which has resolving power of
$\lambda/\Delta\lambda \approx 46,\!000$ and a wavelength coverage of
$\sim$3600--7400\,\AA\@.  The spectra were extracted and analyzed
according to the procedures outlined by \citet{buchhave:2010}.  Having
monitored the IAU radial velocity standard HD182488 over the span of
about 400 days with both TRES and FIES, we can compare the velocity
zero-points and stability of each instrument to determine if any
correction must be applied before combining the datasets.  We
calculate the mean TRES velocity to be $-20.821 \pm 0.060\,\kms$ (rms
error) when correlations are performed in a single echelle order
against a synthetic spectrum.  For FIES, this number is $-20.890 \pm
0.046\,\kms$.  From this, we conclude that velocities from TRES and
FIES are on the same velocity scale (to within the errors), and for
the purposes of detecting velocity variation due to a stellar
companion (tens of \kms), no offset need be applied.

Based on the reconnaissance spectroscopy, we find that the system has
rms velocity residuals consistent with no velocity variation within
the measurement precision, and the observations show no evidence of a
composite spectrum.  From this we conclude that \hatcur{} has no
stellar companion.  Furthermore, the surface gravity found by each
instrument is consistent with a dwarf star, which reduces the
likelihood that the HATNet detection is caused by a background blend.

While we found the reconnaissance velocities to be consistent with
each other to within the measurement precision, it is worth noting
that those velocities were calculated by making use of only one order
of each spectrum, a small wavelength range ($\sim$80\,\AA\@)
surrounding the Mg I b triplet.  By cross-correlating two spectra
against each other order-by-order and summing the correlation
functions, we can utilize the information in a larger wavelength
range, effectively increasing the S/N and reducing the measurement
errors.  In doing so, we may be able to detect a statistically
significant velocity variation, which would be further evidence for a
planetary companion.  Of course, failure to detect such a variation
would not rule out a planet; it would simply place an upper limit on
any companion mass.  For the multi-order cross-correlation, we used
observed templates -- spectra of the IAU standard HD182488 taken on
the same night as each \hatcur{} spectrum -- to perform the analysis.
This allowed us to shift each relative velocity to the IAU scale and
look for orbital motion.  We used the wavelength range
$\sim$4400--6650\,\AA\@, and the resulting velocities are shown in
\reftabl{recon}.  The revised velocities imply
$\mpl=\ensuremath{0.74\pm0.19}\,\mjup$, assuming a mass of 1\,\msun\
for \hatcur{}, which is consistent with the classification from the
reconnaissance spectra.  The error does not take into account
systematic uncertainties due to template mismatch, which could affect
the relative velocities; or the stability of the instruments, which
could affect the combination of the relative velocities onto the
absolute scale.  Nonetheless, this result lends confidence to the
decision to move forward with more precious follow-up resources for
precise velocities.

\ifthenelse{\boolean{emulateapj}}{
    \begin{deluxetable*}{llcccc}
}{
    \begin{deluxetable}{llcccc}
}
\tablewidth{0pc}
\tablecaption{
	Reconnaissance Spectroscopy of \hatcur{}.
	\label{tab:recon}
}
\tablehead{
	\colhead{Intsrument} & 
	\colhead{BJD} & 
	\colhead{RV\tablenotemark{a}} & 
	\colhead{\teffstar\tablenotemark{b}} & 
	\colhead{\loggstar} & 
	\colhead{\vsini}\\
	\colhead{} & 
	\colhead{(2,454,000$+$)} & 
	\colhead{(\kms)} &
	\colhead{(K)} &
	\colhead{} &
	\colhead{(\kms)}
}
\startdata
TRES & $ 1113.87949 $ & $ -12.417 \pm 0.036 $ & $ 5250 \pm 125 $ & $ 4.0 \pm 0.25 $ & $ 6.0 \pm 1.0 $\\
FIES & $ 1115.67533 $ & $ -12.608 \pm 0.036 $ & $ 5250 \pm 125 $ & $ 4.0 \pm 0.25 $ & $ 4.0 \pm 1.0 $\\
   [-1.5ex]
\enddata
\tablenotetext{a}{
	The velocities reported here are the result of a multi-order
    cross-correlation against the IAU standard HD182488, and have been
    shifted to the absolute velocity scale of \citet{nidever:2002}. 
    The errors are calculated from the rms scatter of individual orders
    in the multi-order analysis.  This error estimate does not include
    possible systematic errors due to, e.g., template mismatch or
    instrument stability.
}
\tablenotetext{b}{
	The stellar parameters \teffstar, \loggstar, and \vsini{} are
	the result of correlation of a single order against a grid of
	synthetic spectra with assumed $\feh=0$.  Errors quoted are
	half of the grid spacing, but do not take into account
	systematic errors introduced if \hatcur{} has non-solar
	metallicity.
}
\ifthenelse{\boolean{emulateapj}}{
    \end{deluxetable*}
}{
    \end{deluxetable}
}

\subsection{High resolution, high S/N spectroscopy}
\label{sec:hispec}

Given the significant transit detection by HATNet, and the encouraging
results from FIES and TRES that rule out obvious false positives, we
proceeded with the follow-up of this candidate by obtaining
high-resolution, high-S/N spectra to characterize the RV variations,
and to refine the determination of the stellar parameters.  For this
we used the HIRES instrument \citep{vogt:1994} on the Keck~I telescope
located on Mauna Kea, Hawaii, between 2009 December and 2010 February.
The width of the spectrometer slit was $0\farcs86$, resulting in a
resolving power of $\lambda/\Delta\lambda \approx 55,\!000$, with a
wavelength coverage of $\sim$3800--8000\,\AA\@.

\begin{figure}[ht]
\ifthenelse{\boolean{emulateapj}}{
	\plotone{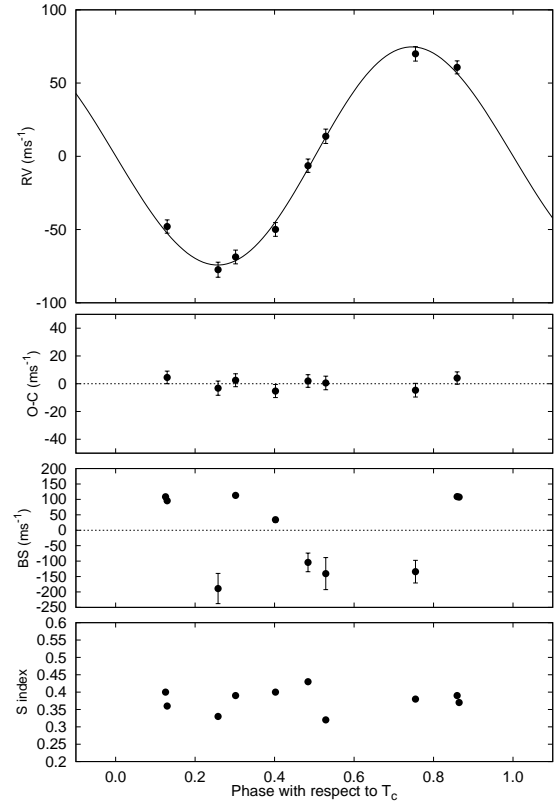}
}{
	\includegraphics[scale=0.8]{\hatcurhtr-rv.eps}
}
\caption{
	{\em Top panel:} Keck/HIRES RV measurements for
        \hbox{\hatcur{}} shown as a function of orbital phase, along
        with our best-fit model (see \reftabl{planetparam}).  Zero
        phase corresponds to the time of mid-transit.  The
        center-of-mass velocity has been subtracted.
	{\em Second panel:} Velocity $O\!-\!C$ residuals from the best
        fit. The error bars include a component from astrophysical
        jitter ($\hatcurRVjitter$\,\ms) added in quadrature to the
        formal errors (see \refsecl{globmod}).
	{\em Third panel:} Bisector spans (BS), with the mean value
        subtracted.  The measurement from the template spectrum is
        included (see \refsecl{bisec}).
	{\em Bottom panel:} Relative chromospheric activity index $S$
        measured from the Keck spectra.
	Note the different vertical scales of the panels.
\label{fig:rvbis}}
\end{figure}

We obtained 8 exposures through an iodine gas absorption cell, which
was used to superimpose a dense forest of $\mathrm{I}_2$ lines on the
stellar spectrum and establish an accurate wavelength fiducial
\citep[see][]{marcy:1992}.  An additional two exposures were taken
without the iodine cell, for use as a template in the reductions.
Relative RVs in the solar system barycentric frame were derived as
described by \cite{butler:1996}, incorporating full modeling of the
spatial and temporal variations of the instrumental profile.  The RV
measurements and their uncertainties are listed in \reftabl{rvs}.  The
period-folded data, along with our best fit described below in
\refsecl{analysis}, are displayed in \reffigl{rvbis}.

In the same figure we show also the relative $S$ index, which is a
measure of the chromospheric activity of the star derived from the
flux in the cores of the \ion{Ca}{2} H and K lines.  This index was
computed following the prescription given by \citet{vaughan:1978}, but
was not calibrated to their scale (for details on calculation of the
relative S index, see \citealt{hartman:2009}).  We do not detect any
significant variation of the relative S index correlated with orbital
phase; such a correlation might have indicated that the RV variations
could be due to stellar activity, casting doubt on the planetary
nature of the candidate.  There is no sign of emission in the cores of
the \ion{Ca}{2} H and K lines in any of our spectra, from which we
conclude that the chromospheric activity in \hatcur{} is very low.
Furthermore, we estimate from the HIRES spectra that on the Mt.
Wilson activity scale adopted by \citet{knutson:2010}, the $\logrhk =
\hatcurlogrhk$ and $S_{HK} = \hatcurshk$ (median values).  The
$\logrhk$ value depends on our estimate of $B-V = 0.75$ using the
\teff{} calibration of \citet{valenti:2005}.

\ifthenelse{\boolean{emulateapj}}{
    \begin{deluxetable*}{lrrrrr}
}{
    \begin{deluxetable}{lrrrrr}
}
\tablewidth{0pc}
\tablecaption{
	Relative radial velocities, bisector spans, and activity index
	measurements of \hatcur{}.
	\label{tab:rvs}
}
\tablehead{
	\colhead{BJD} & 
	\colhead{RV\tablenotemark{a}} & 
	\colhead{\ensuremath{\sigma_{\rm RV}}\tablenotemark{b}} & 
	\colhead{BS} & 
	\colhead{\ensuremath{\sigma_{\rm BS}}} & 
	\colhead{S\tablenotemark{c}}\\
	\colhead{\hbox{(2,454,000$+$)}} & 
	\colhead{(\ms)} & 
	\colhead{(\ms)} &
	\colhead{(\ms)} &
    \colhead{(\ms)} &
	\colhead{}
}
\startdata
$ 1188.91303 $ & $   -68.72 $ & $     2.58 $ & $   113.04 $ & $     4.61 $ & $    0.39 $ \\
$ 1190.94964 $ & $    60.74 $ & $     2.14 $ & $   109.06 $ & $     4.41 $ & $    0.39 $ \\
$ 1190.96797 $ & \nodata      & \nodata      & $   107.63 $ & $     4.51 $ & $    0.37 $ \\
$ 1191.92204 $ & \nodata      & \nodata      & $   108.46 $ & $     7.58 $ & $    0.40 $ \\
$ 1191.93647 $ & $   -47.94 $ & $     2.31 $ & $    95.72 $ & $     4.43 $ & $    0.36 $ \\
$ 1192.93217 $ & $   -49.91 $ & $     2.62 $ & $    34.02 $ & $     7.53 $ & $    0.40 $ \\
$ 1196.88472 $ & $    -6.40 $ & $     2.36 $ & $  -104.19 $ & $    30.34 $ & $    0.43 $ \\
$ 1197.87232 $ & $    69.96 $ & $     3.01 $ & $  -134.17 $ & $    36.62 $ & $    0.38 $ \\
$ 1250.85004 $ & $   -77.40 $ & $     3.34 $ & $  -188.93 $ & $    48.96 $ & $    0.33 $ \\
$ 1251.84005 $ & $    13.68 $ & $     2.98 $ & $  -140.64 $ & $    51.95 $ & $    0.32 $ \\
        [-1.5ex]
\enddata
\tablenotetext{a}{
	The zero-point of these velocities is arbitrary. An overall offset
    $\gamma_{\rm rel}$ fitted to these velocities in \refsecl{globmod}
    has {\em not} been subtracted.
}
\tablenotetext{b}{
	Internal errors excluding the component of astrophysical jitter
    considered in \refsecl{globmod}.
}
\tablenotetext{c}{
	Relative chromospheric activity index, not calibrated to the scale
	of \citet{vaughan:1978}.
}
\tablecomments{
	Note that for the iodine-free template exposures we do not
        measure the RV but do measure the BS and S index.  Such
        template exposures can be distinguished by the missing RV
        value.
}
\ifthenelse{\boolean{emulateapj}}{
    \end{deluxetable*}
}{
    \end{deluxetable}
}

\subsection{Photometric follow-up observations}
\label{sec:phot}

\begin{figure}[!ht]
\plotone{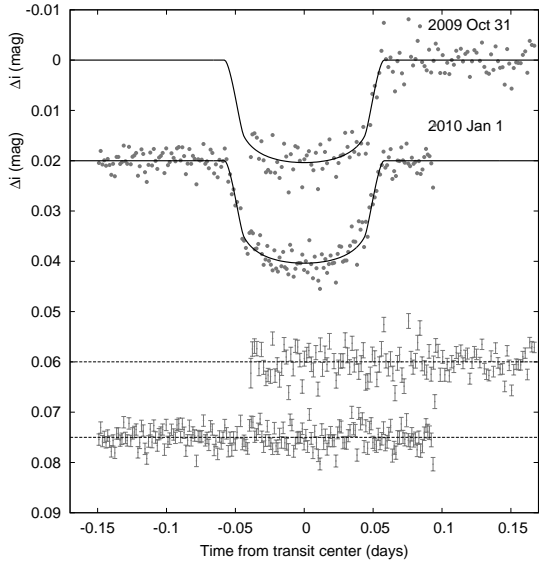}
\caption{
	Unbinned instrumental \band{i} transit \lcs{}, acquired with
    KeplerCam at the \flwof{} telescope on 2009 Oct 31 and 2010 Jan 1. 
    The light curves have been EPD- and TFA-processed, as described in
    \refsec{globmod}.  Curves after the first are displaced vertically
    for clarity.  Our best fit from the global modeling described in
    \refsecl{globmod} is shown by the solid lines.  Residuals from the
    fits are displayed at the bottom, in the same order as the top
    curves.  The error bars represent the photon and background shot
    noise, plus the readout noise.
\label{fig:lc}}
\end{figure}

In order to permit a more accurate modeling of the light curve, we
conducted additional photometric observations with the KeplerCam CCD
camera on the \flwof{} telescope.  We observed two transit events of
\hatcur{} on the nights of 2009 Oct 31 and 2010 Jan 1 (\reffigl{lc}).
On 2009 Oct 31, $124$ frames were acquired with a cadence of $133$
seconds ($120$ seconds of exposure time) in the Sloan \band{i}, while
on 2010 Jan 1, $190$ frames were acquired with a cadence of $109$
seconds ($80$ seconds of exposure time) in the Sloan \band{i}.

The reduction of these images, including basic calibration,
astrometry, and aperture photometry, was performed as described by
\citet{bakos:2009}.  We performed EPD and TFA to remove trends
simultaneously with the light curve modeling (for more details, see
\refsecl{analysis}, and \citealt{bakos:2009}).  The final time series
are shown in the top portion of \reffigl{lc}, along with our best-fit
transit \lc{} model described below; the individual measurements are
reported in \reftabl{phfu}.

\begin{deluxetable}{lrrrr}
\tablewidth{0pc}
\tablecaption{High-precision differential photometry of
        \hatcur\label{tab:phfu}}
\tablehead{
	\colhead{BJD} & 
	\colhead{Mag\tablenotemark{a}} & 
	\colhead{\ensuremath{\sigma_{\rm Mag}}} &
	\colhead{Mag(orig)\tablenotemark{b}} & 
	\colhead{Filter} \\
	\colhead{\hbox{~~~~(2,400,000$+$)~~~~}} & 
	\colhead{} & 
	\colhead{} &
	\colhead{} & 
	\colhead{}
}
\startdata
$ 55136.63175 $ & $   0.01928 $ & $   0.00295 $ & $  11.52150 $ & $ i$\\
$ 55136.63284 $ & $   0.01476 $ & $   0.00144 $ & $  11.52080 $ & $ i$\\
$ 55136.63437 $ & $   0.01455 $ & $   0.00145 $ & $  11.51890 $ & $ i$\\
$ 55136.63608 $ & $   0.01541 $ & $   0.00144 $ & $  11.52220 $ & $ i$\\
$ 55136.63764 $ & $   0.01963 $ & $   0.00142 $ & $  11.52500 $ & $ i$\\
$ 55136.63939 $ & $   0.01613 $ & $   0.00142 $ & $  11.52420 $ & $ i$\\
$ 55136.64095 $ & $   0.01992 $ & $   0.00139 $ & $  11.52700 $ & $ i$\\
$ 55136.64269 $ & $   0.02231 $ & $   0.00145 $ & $  11.53230 $ & $ i$\\
$ 55136.64424 $ & $   0.02184 $ & $   0.00133 $ & $  11.53140 $ & $ i$\\
$ 55136.64600 $ & $   0.02230 $ & $   0.00136 $ & $  11.53070 $ & $ i$\\
        [-1.5ex]
\enddata
\tablenotetext{a}{
	The out-of-transit level has been subtracted. These magnitudes have
        been subjected to the EPD and TFA procedures, carried out
        simultaneously with the transit fit.
}
\tablenotetext{b}{
	Raw magnitude values without application of the EPD and TFA
	procedures.
}
\tablecomments{
    This table is available in a machine-readable form in the online
    journal.  A portion is shown here for guidance regarding its form
    and content.
}
\end{deluxetable}

\section{Analysis}
\label{sec:analysis}

\subsection{Properties of the parent star}
\label{sec:stelparam}

Fundamental parameters of the host star \hatcur{} such as the mass
(\mstar) and radius (\rstar), which are needed to infer the planetary
properties, depend strongly on other stellar quantities that can be
derived spectroscopically.  For this we have relied on our template
spectrum obtained with the Keck/HIRES instrument, and the analysis
package known as Spectroscopy Made Easy \citep[SME;][]{valenti:1996},
along with the atomic line database of \cite{valenti:2005}.  SME
yielded the following {\em initial} values and uncertainties:
effective temperature $\teffstar=\hatcurSMEiteff$\,K, 
stellar surface gravity $\loggstar=\hatcurSMEilogg$\,(cgs),
metallicity $\feh=\hatcurSMEizfeh$, and 
projected rotational velocity $\vsini=\hatcurSMEivsin\,\kms$.

In principle the stellar effective temperature and metallicity, along
with the stellar surface gravity taken as a luminosity indicator,
could be used as constraints to infer the stellar mass and radius by
comparison with stellar evolution models.  However, because of reasons
described in \citet{sozzetti:2007}, we used the \arstar\ normalized
semi-major axis as luminosity indicator instead of \loggstar.  The
\arstar\ quantity is closely related to \rhostar, the mean stellar
density, and can be derived directly from the transit \lcs\
\citep{sozzetti:2007} and the RV data (for eccentric cases, see
\refsecl{globmod}).  This, in turn, allows us to improve on the
determination of the spectroscopic parameters by supplying an indirect
constraint on the weakly determined spectroscopic value of \loggstar\
that removes degeneracies.  We take this approach here, as described
below.  The validity of our assumption, namely that the adequate
physical model describing our data is a planetary transit (as opposed
to a blend), is shown later in \refsecl{bisec}.

Our initial values of \teffstar, \loggstar, and \feh\ were used to
determine auxiliary quantities needed in the global modeling of the
follow-up photometry and radial velocities (specifically, the
limb-darkening coefficients).  This modeling, the details of which are
described in \refsecl{globmod}, uses a Monte Carlo approach to deliver
the numerical probability distribution of \arstar\ and other fitted
variables.  For further details we refer the reader to
\cite{pal:2009b}.  When combining \arstar\ (used as a proxy for
luminosity) with assumed Gaussian distributions for \teffstar\ and
\feh\ based on the SME determinations, a comparison with stellar
evolution models allows the probability distributions of other stellar
properties to be inferred, including \loggstar.  Here we use the
stellar evolution calculations from the \hatcurisofull\ series by
\cite{\hatcurisocite}.  The comparison against the model isochrones
was carried out for each of 20,000 Monte Carlo trial sets (see
\refsecl{globmod}).  Parameter combinations corresponding to
unphysical locations in the \hbox{H-R} diagram (53\% of the trials)
were ignored, and replaced with another randomly drawn parameter set.
The result for the surface gravity, $\loggstar = \hatcurISOlogg$, is
not significantly different from our initial SME analysis.  However,
we carried out a second iteration anyway, in which we adopted this
value of \loggstar\ and held it fixed in a new SME analysis (coupled
with a new global modeling of the RV and \lcs), adjusting only
\teffstar, \feh, and \vsini.  This gave
$\teffstar = \hatcurSMEiiteff$\,K, 
$\feh = \hatcurSMEiizfeh$, and 
$\vsini = \hatcurSMEiivsin$\,\kms,
in which the uncertainties for the first two have been increased by a
factor of two over their formal values to include our estimates of the
systematic uncertainties.  A further iteration did not change
\loggstar\ significantly, so we adopted the values stated above as the
final atmospheric properties of the star.  They are collected in
\reftabl{stellar}, together with the adopted values for the
macroturbulent and microturbulent velocities.

\begin{figure}[!ht]
\plotone{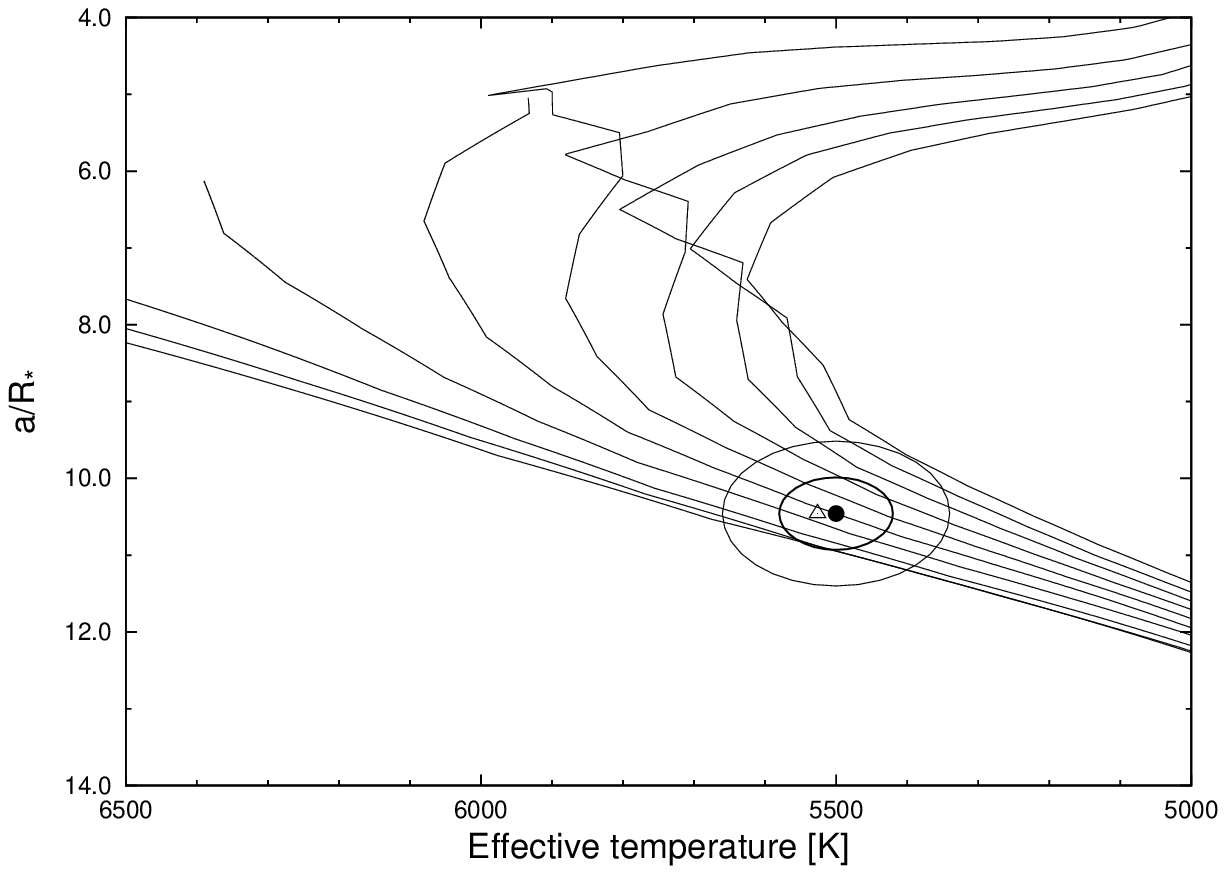}
\caption{
	Model isochrones from \cite{\hatcurisocite} for the measured
        metallicity of \hatcur, \feh = \hatcurSMEiizfehshort, and ages
        of 0.2, 0.5, 1.0, 2.0, 3.0, 4.0, 5.0, 6.0, 7.0, and 8.0\,Gyr
        (left to right).  The adopted values of $\teffstar$ and
        \arstar\ are shown together with their 1$\sigma$ and 2$\sigma$
        confidence ellipsoids. The initial values of \teffstar\ and
        \arstar\ from the first SME and \lc\ analyses are represented
        with a triangle.
\label{fig:iso}}
\end{figure}

With the adopted spectroscopic parameters the model isochrones yield
the stellar mass and radius \mstar\ = \hatcurISOmlong\,\msun\ and
\rstar\ = \hatcurISOrlong\,\rsun, along with other properties listed
at the bottom of \reftabl{stellar}.  \hatcur{} is a \hatcurISOspec\
dwarf star with an estimated age of \hatcurISOage\,Gyr, according to
these models.  The inferred location of the star in a diagram of
\arstar\ versus \teffstar, analogous to the classical H-R diagram, is
shown in \reffigl{iso}.  The stellar properties and their 1$\sigma$
and 2$\sigma$ confidence ellipsoids are displayed against the backdrop
of \cite{\hatcurisocite} isochrones for the measured metallicity of
\feh\ = \hatcurSMEiizfehshort, and a range of ages.  For comparison,
the location implied by the initial SME results is also shown
(triangle).

The stellar evolution modeling provides color indices that may be
compared against the measured values as a sanity check.  The best
available measurements are the near-infrared magnitudes from the 2MASS
Catalogue \citep{skrutskie:2006},
$J_{\rm 2MASS}=\hatcurCCtwomassJmag$, 
$H_{\rm 2MASS}=\hatcurCCtwomassHmag$, and 
$K_{\rm 2MASS}=\hatcurCCtwomassKmag$;
which we have converted to the photometric system of the models (ESO
system) using the transformations by \citet{carpenter:2001}.  The
resulting measured color index is $J-K = \hatcurCCesoJKmag$.  This
value is somewhat higher than the value of $J-K = \hatcurISOJK$
predicted from the isochrones, and suggests that the star may be
affected by interstellar reddening.  Estimates of the total reddening
along the line of sight may be obtained from the dust maps by
\citet{schlegel:1998}\footnote{see
http://irsa.ipac.caltech.edu/applications/DUST} and
\citet{burstein:1982}.  The average of these two values is $E(B\!-\!V)
= \hatcurEBVtotal$.  The fraction of this reddening amount that
applies to \hatcur{} depends on the distance to the object and its
Galactic latitude \citep[see, e.g.][]{bonifacio:2000}.  For the
distance we use an initial estimate derived from the absolute $K$-band
magnitude predicted by the models ($M_{\rm K}=\hatcurISOMK$) and the
apparent magnitude in the $K_s$ band from 2MASS, which is less
affected by extinction, and which we convert to the system of the
isochrones as before.  The process is iterated, and results in a final
reddening of $E(B\!-\!V) = \hatcurEBV$, a de-reddened color of $J-K =
\hatcurCCesoJKmagf$ in good agreement with the models, and a final
distance of $\hatcurXdist$\,pc.  These values are listed in
\reftabl{stellar}.

\begin{deluxetable}{lrl}
\tablewidth{0pc}
\ifthenelse{\boolean{emulateapj}}{
	\tabletypesize{\scriptsize}
}{
	\tabletypesize{\footnotesize}
}
\tablecaption{
	Stellar parameters for \hatcur{}
	\label{tab:stellar}
}
\tablehead{
	\colhead{~~~~~~~~Parameter~~~~~~~~}	&
	\colhead{Value}                         &
	\colhead{Source}
}
\startdata
\noalign{\vskip -3pt}
\sidehead{Spectroscopic properties}
~~~~$\teffstar$ (K)\dotfill         &  \hatcurSMEteff       & SME\tablenotemark{a}\\
~~~~$\feh$ \dotfill                 &  \hatcurSMEzfeh       & SME                 \\
~~~~$\vsini$ (\kms)\dotfill         &  \hatcurSMEvsin       & SME                 \\
~~~~$\vmac$ (\kms)\dotfill          &  \hatcurSMEvmac       & SME                 \\
~~~~$\vmic$ (\kms)\dotfill          &  \hatcurSMEvmic       & SME                 \\
~~~~$\logrhk$ \dotfill              &  \hatcurlogrhk        & HIRES\tablenotemark{b}          \\
~~~~$\gamma_{\rm RV}$ (\kms)\dotfill &  \hatcurDSgamma      & TRES+FIES\tablenotemark{c}          \\
\sidehead{Photometric properties}
~~~~$V$ (mag)\dotfill               &  \hatcurCCtassmv      & TASS            \\
~~~~$V\!-\!I_C$ (mag)\dotfill       &  \hatcurCCtassvi      & TASS            \\
~~~~$J$ (mag)\dotfill               &  \hatcurCCtwomassJmag & 2MASS           \\
~~~~$H$ (mag)\dotfill               &  \hatcurCCtwomassHmag & 2MASS           \\
~~~~$K_s$ (mag)\dotfill             &  \hatcurCCtwomassKmag & 2MASS           \\
\sidehead{Derived properties}
~~~~$\mstar$ ($\msun$)\dotfill      &  \hatcurISOmlong      & \hatcurisoshort+\hatcurlumind+SME \tablenotemark{d}\\
~~~~$\rstar$ ($\rsun$)\dotfill      &  \hatcurISOrlong      & \hatcurisoshort+\hatcurlumind+SME         \\
~~~~$\loggstar$ (cgs)\dotfill       &  \hatcurISOlogg       & \hatcurisoshort+\hatcurlumind+SME         \\
~~~~$\lstar$ ($\lsun$)\dotfill      &  \hatcurISOlum        & \hatcurisoshort+\hatcurlumind+SME         \\
~~~~$M_V$ (mag)\dotfill             &  \hatcurISOmv         & \hatcurisoshort+\hatcurlumind+SME         \\
~~~~$M_K$ (mag,\hatcurjhkfilset)\dotfill &  \hatcurISOMK    & \hatcurisoshort+\hatcurlumind+SME         \\
~~~~Age (Gyr)\dotfill               &  \hatcurISOage        & \hatcurisoshort+\hatcurlumind+SME         \\
~~~~$E(B\!-\!V)$ (mag)\dotfill      &  \hatcurEBV           & 2MASS+\hatcurisoshort+\hatcurlumind+SME\tablenotemark{e}\\
~~~~Distance (pc)\dotfill           &  \hatcurXdist         & 2MASS+\hatcurisoshort+\hatcurlumind+SME   \\
[-1.5ex]
\enddata
\tablenotetext{a}{
	SME = ``Spectroscopy Made Easy'' package for the analysis of
        high-resolution spectra \citep{valenti:1996}. These parameters
        rely primarily on SME, but have a small dependence also on the
        iterative analysis incorporating the isochrone search and
        global modeling of the data, as described in the text.
}
\tablenotetext{b}{
	See \refsecl{hispec}.
}
\tablenotetext{c}{
	This velocity is on the absolute scale of \citet{nidever:2002}.  We
	have increased the error to include to include the rms error of
	individual orders in the multi-order correlation as well as an
	estimate of possible systematic errors introduced by the choice of
	template spectrum and the stability of the instruments.
}
\tablenotetext{d}{
	\hatcurisoshort+\hatcurlumind+SME = Based on the 
        \hatcurisoshort\ isochrones \citep{\hatcurisocite},
        \hatcurlumind\ as a luminosity indicator, and the SME results.
}
\tablenotetext{e}{
	The {\em total} $E(B\!-\!V)$ along the line of sight comes from
	\citet{schlegel:1998} and \citet{burstein:1982}.  The final
	distance and $E(B\!-\!V)$ at that distance are calculated
	iteratively using $M_K$ and $K_s$ (see \refsecl{stelparam}).
}
\end{deluxetable}

\subsection{Spectral line-bisector analysis}
\label{sec:bisec}

Our initial spectroscopic analyses discussed in \refsecl{recspec} and
\refsecl{hispec} rule out the most obvious astrophysical false positive
scenarios.  However, more subtle phenomena such as blends
(contamination by an unresolved eclipsing binary, whether in the
background or associated with the target) can still mimic both the
photometric and spectroscopic signatures we see.

Following \cite{torres:2007}, we explored the possibility that the
measured radial velocities are not real, but are instead caused by
distortions in the spectral line profiles due to contamination from a
nearby unresolved eclipsing binary.  A bisector analysis based on the
Keck spectra was done as described in \S 5 of \cite{bakos:2007a}.

\begin{figure}[!ht]
\ifthenelse{\boolean{emulateapj}}{
	\plotone{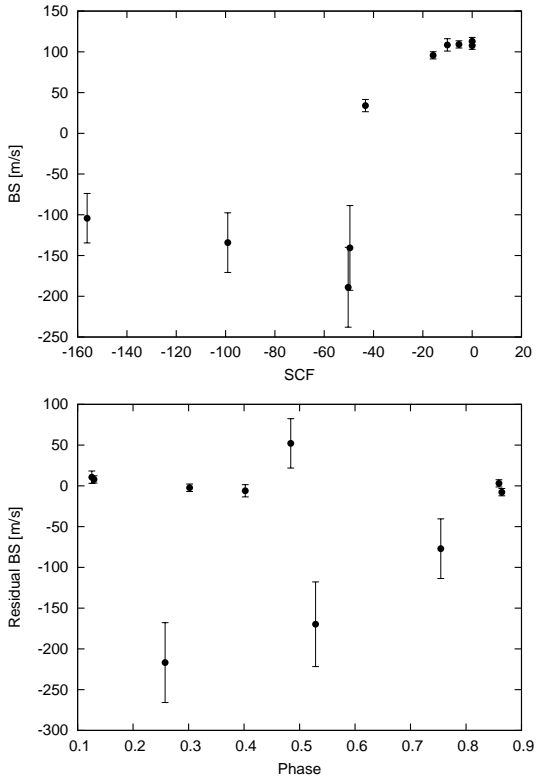}
}{
	\includegraphics[scale=0.8]{\hatcurhtr-SCF.eps}
}
\caption{
        {\em Top panel:} The expected BS variation due to sky
        contamination (SCF) compared to the observed variation.  The
        correlation between these values indicates that the observed
        BS variation for this relatively faint target appears to be
        due to contamination from scattered moonlight.  {\em Bottom
        panel:} The residual BS variation when corrected for the SCF,
        compared to orbital phase.  See \refsec{bisec}.
\label{fig:scf}}
\end{figure}

The bisector spans (BS) show significant variations, however they do
not correlate with the stellar RV as would be expected if the apparent
RV variations were due to a blend.  Following our earlier work
\citep{kovacs:2010,hartman:2009} we investigated the effect of
contamination from moonlight on the measured bisector spans (BS).  By
adopting the same definition of the sky contamination factor SCF as in
\citet{kovacs:2010}, we find a correlation between the SCF and BS (see
\reffigl{scf}, top panel).  When we correct for the relation
(\reffigl{scf}, bottom panel), the residual BS show no correlation
with orbital phase.  There do appear to be a few discrepant BS
residuals, although this is the faintest star for which we have
performed this analysis, and the sky contamination appears to be
especially pernicious in this case, both of which contribute
uncertainties that are not fully characterized.  As expected, the
outliers correspond to the spectra with the lowest S/N and strongest
SCF.  We conclude that the observed BS variations are due to
contamination from scattered moonlight, that the velocity variations
are real, and that the star is orbited by a close-in giant planet.

\subsection{Global modeling of the data}
\label{sec:globmod}
This section describes the procedure we followed to model the HATNet
photometry, the follow-up photometry, and the radial velocities
simultaneously.  Our model for the follow-up \lcs\ used analytic
formulae based on \citet{mandel:2002} for the eclipse of a star by a
planet, with limb darkening being prescribed by a quadratic law.  The
limb darkening coefficients for the Sloan \band{i} were interpolated
from the tables by \citet{claret:2004} for the spectroscopic
parameters of the star as determined from the SME analysis
(\refsecl{stelparam}).  The transit shape was parametrized by the
normalized planetary radius $p\equiv \rpl/\rstar$, the square of the
impact parameter $b^2$, and the reciprocal of the half duration of the
transit $\zrstar$.  We chose these parameters because of their simple
geometric meanings and the fact that these show negligible
correlations \citep[see][]{bakos:2009,kipping:2010}.  Our model for
the HATNet data was the simplified ``P1P3'' version of the
\citet{mandel:2002} analytic functions (an expansion in terms of
Legendre polynomials), for the reasons described in
\citet{bakos:2009}.  Following the formalism presented by
\citet{pal:2009}, the RVs were fitted with an eccentric Keplerian
model parametrized by the semi-amplitude $K$ and Lagrangian elements
$k \equiv e \cos\omega$ and $h \equiv e \sin\omega$, in which $\omega$
is the longitude of periastron.

We assumed that there is a strict periodicity in the individual
transit times.  We assigned the transit number $N_{tr} = 0$ to the
first complete follow-up \lc\ gathered on 2010 Jan 01.  The adjustable
parameters in the fit that determine the ephemeris were chosen to be
the time of the first transit center observed with HATNet,
$T_{c,-129}$, and that of the last transit center observed with the
\flwof\ telescope, $T_{c,0}$.  We used these as opposed to period and
reference epoch in order to minimize correlations between parameters
\citep[see][]{pal:2008}.  Times of mid-transit for intermediate events
were interpolated using these two epochs and the corresponding transit
number of each event, $N_{tr}$.  The eight main parameters describing
the physical model were thus $T_{c,-129}$, $T_{c,0}$, $\rpl/\rstar$,
$b^2$, $\zrstar$, $K$, $k \equiv e\cos\omega$, and $h \equiv
e\sin\omega$.  Three additional parameters were included that have to
do with the instrumental configuration.  These are the HATNet blend
factor $B_{\rm inst}$, which accounts for possible dilution of the
transit in the HATNet \lc\ from background stars due to the broad PSF
(20\arcsec\ FWHM), the HATNet out-of-transit magnitude $M_{\rm
0,HATNet}$, and the relative zero-point $\gamma_{\rm rel}$ of the Keck
RVs.

We extended our physical model with an instrumental model that
describes brightness variations caused by systematic errors in the
measurements.  This was done in a similar fashion to the analysis
presented by \citet{bakos:2009}.  The HATNet photometry has already
been EPD- and TFA-corrected before the global modeling, so we only
considered corrections for systematics in the follow-up \lcs.  We
chose the ``ELTG'' method, i.e., EPD was performed in ``local'' mode
with EPD coefficients defined for each night, and TFA was performed in
``global'' mode using the same set of stars and TFA coefficients for
all nights.  The five EPD parameters were the hour angle (representing
a monotonic trend that changes linearly over time), the square of the
hour angle (reflecting elevation), and the stellar profile parameters
(equivalent to FWHM, elongation, and position angle of the image).
The functional forms of the above parameters contained six
coefficients, including the auxiliary out-of-transit magnitude of the
individual events.  The EPD parameters were independent for both
nights, implying 12 additional coefficients in the global fit.  For
the global TFA analysis we chose 20 template stars that had good
quality measurements for all nights and on all frames, implying an
additional 20 parameters in the fit.  Thus, the total number of fitted
parameters was 11 (physical model with 3 configuration-related
parameters) + 12 (local EPD) + 20 (global TFA) = 43, i.e.~much smaller
than the number of data points (322, counting only RV measurements and
follow-up photometry measurements).

The joint fit was performed as described in \citet{bakos:2009}.  We
minimized \chisq\ in the space of parameters by using a hybrid
algorithm, combining the downhill simplex method \citep[AMOEBA;
see][]{press:1992} with a classical linear least squares algorithm.
Uncertainties for the parameters were derived applying the Markov
Chain Monte-Carlo method \citep[MCMC, see][]{ford:2006} using
``Hyperplane-CLLS'' chains \citep{bakos:2009}.  This provided the full
{\em a posteriori} probability distributions of all adjusted
variables.  The {\em a priori} distributions of the parameters for
these chains were chosen to be Gaussian, with eigenvalues and
eigenvectors derived from the Fisher covariance matrix for the
best-fit solution.  The Fisher covariance matrix was calculated
analytically using the partial derivatives given by \citet{pal:2009}.

Following this procedure we obtained the {\em a posteriori}
distributions for all fitted variables, and other quantities of
interest such as \arstar. As described in \refsecl{stelparam},
\arstar\ was used together with stellar evolution models to infer a
theoretical value for \loggstar\ that is significantly more accurate
than the spectroscopic value. The improved estimate was in turn
applied to a second iteration of the SME analysis, as explained
previously, in order to obtain better estimates of \teffstar\ and
\feh.  The global modeling was then repeated with updated
limb-darkening coefficients based on those new spectroscopic
determinations. The resulting geometric parameters pertaining to the
light curves and velocity curves are listed in \reftabl{planetparam}.

Included in this table is the RV ``jitter''. This is a component of
assumed astrophysical noise intrinsic to the star that we added in
quadrature to the internal errors for the RVs in order to achieve
$\chi^{2}/{\rm dof} = 1$ from the RV data for the global fit.
Auxiliary parameters not listed in the table are:
$T_{\mathrm{c},-129}=\hatcurLCTA$~(BJD),
$T_{\mathrm{c},0}=\hatcurLCTB$~(BJD), the blending factor
$B_{\rm instr}=\hatcurLCiblend$, and
$\gamma_{\rm rel}=\hatcurRVgamma$\,\ms. 
The latter quantity represents an arbitrary offset for the Keck RVs,
and does \emph{not} correspond to the true center of mass velocity of
the system, which was listed earlier in \reftabl{stellar}
($\gamma_{\rm RV}$).

The planetary parameters and their uncertainties can be derived by
combining the {\em a posteriori} distributions for the stellar, \lc,
and RV parameters.  In this way we find a mass for the planet of
$\mpl=\hatcurPPmlong\,\mjup$ and a radius of
$\rpl=\hatcurPPrlong\,\rjup$, leading to a mean density
$\rho_p=\hatcurPPrho$\,\gcmc. 
These and other planetary parameters are listed at the bottom of
Table~\ref{tab:planetparam}.  We note that the eccentricity of the
orbit is not significant: $k = \hatcurRVk$, $h = \hatcurRVh$ ($e =
\hatcurRVeccen$, $\omega = \hatcurRVomega\arcdeg$).

\begin{deluxetable}{lr}
\tabletypesize{\scriptsize}
\tablecaption{Orbital and planetary parameters\label{tab:planetparam}}
\tablehead{
	\colhead{~~~~~~~~~~~~~~~Parameter~~~~~~~~~~~~~~~} &
	\colhead{Value}
}
\startdata
\noalign{\vskip -3pt}
\sidehead{\Lc{} parameters}
~~~$P$ (days)             \dotfill    & $\hatcurLCP$              \\
~~~$T_c$ (${\rm BJD}$)    
      \tablenotemark{a}   \dotfill    & $\hatcurLCT$              \\
~~~$T_{14}$ (days)
      \tablenotemark{a}   \dotfill    & $\hatcurLCdur$            \\
~~~$T_{12} = T_{34}$ (days)
      \tablenotemark{a}   \dotfill    & $\hatcurLCingdur$         \\
~~~$\arstar$              \dotfill    & $\hatcurPPar$             \\
~~~$\zrstar$              \dotfill    & $\hatcurLCzeta$           \\
~~~$\rpl/\rstar$          \dotfill    & $\hatcurLCrprstar$        \\
~~~$b^2$                  \dotfill    & $\hatcurLCbsq$            \\
~~~$b \equiv a \cos i/\rstar$
                          \dotfill    & $\hatcurLCimp$            \\
~~~$i$ (deg)              \dotfill    & $\hatcurPPi$              \\

\sidehead{Limb-darkening coefficients \tablenotemark{b}}
~~~$a_i$ (linear term)    \dotfill    & $\hatcurLBii$             \\
~~~$b_i$ (quadratic term) \dotfill    & $\hatcurLBiii$            \\

\sidehead{RV parameters}
~~~$K$ (\ms)              \dotfill    & $\hatcurRVK$            \\
~~~$k_{\rm RV}$\tablenotemark{c} 
                          \dotfill    & $\hatcurRVk$             \\
~~~$h_{\rm RV}$\tablenotemark{c}
                          \dotfill    & $\hatcurRVh$              \\
~~~$e$                    \dotfill    & $\hatcurRVeccen$          \\
~~~$\omega$ (deg)         \dotfill    & $\hatcurRVomega$          \\
~~~RV jitter (\ms)        \dotfill    & \hatcurRVjitter           \\

\sidehead{Secondary eclipse parameters}
~~~$T_s$ (BJD)            \dotfill    & $\hatcurXsecondary$       \\
~~~$T_{s,14}$             \dotfill    & $\hatcurXsecdur$          \\
~~~$T_{s,12}$             \dotfill    & $\hatcurXsecingdur$       \\

\sidehead{Planetary parameters}
~~~$\mpl$ ($\mjup$)       \dotfill    & $\hatcurPPmlong$          \\
~~~$\rpl$ ($\rjup$)       \dotfill    & $\hatcurPPrlong$          \\
~~~$C(\mpl,\rpl)$
    \tablenotemark{d}     \dotfill    & $\hatcurPPmrcorr$         \\
~~~$\rhopl$ (\gcmc)       \dotfill    & $\hatcurPPrho$            \\
~~~$\log g_p$ (cgs)       \dotfill    & $\hatcurPPlogg$           \\
~~~$a$ (AU)               \dotfill    & $\hatcurPParel$           \\
~~~$T_{\rm eq}$ (K)       \dotfill    & $\hatcurPPteff$           \\
~~~$\Theta$\tablenotemark{e} \dotfill & $\hatcurPPtheta$          \\
~~~$\langle F \rangle$ ($10^{\hatcurPPfluxavgdim}$\ergscmsq) \tablenotemark{f}
                          \dotfill    & $\hatcurPPfluxavg$        \\
[-1.5ex]
\enddata
\tablenotetext{a}{
    \ensuremath{T_c}: Reference epoch of mid transit that
        minimizes the correlation with the orbital period. It
        corresponds to $N_{tr} = -6$. BJD is calculated from UTC.
	\ensuremath{T_{14}}: total transit duration, time
	between first to last contact;
	\ensuremath{T_{12}=T_{34}}: ingress/egress time, time between
        first and second, or third and fourth contact.
}
\tablenotetext{b}{
	Values for a quadratic law, adopted from the tabulations by
        \cite{claret:2004} according to the spectroscopic (SME)
        parameters listed in \reftabl{stellar}.
}
\tablenotetext{c}{
    Lagrangian orbital parameters derived from the global modeling, 
    and primarily determined by the RV data. 
}
\tablenotetext{d}{
	Correlation coefficient between the planetary mass \mpl\ and
        radius \rpl.
}
\tablenotetext{e}{
	The Safronov number is given by $\Theta = \frac{1}{2}(V_{\rm
	esc}/V_{\rm orb})^2 = (a/\rpl)(\mpl / \mstar )$
	\citep[see][]{hansen:2007}.
}
\tablenotetext{f}{
	Incoming flux per unit surface area, averaged over the orbit.
}
\end{deluxetable}


\section{Discussion}
\label{sec:discussion}

\begin{figure*}[!ht]
\plotone{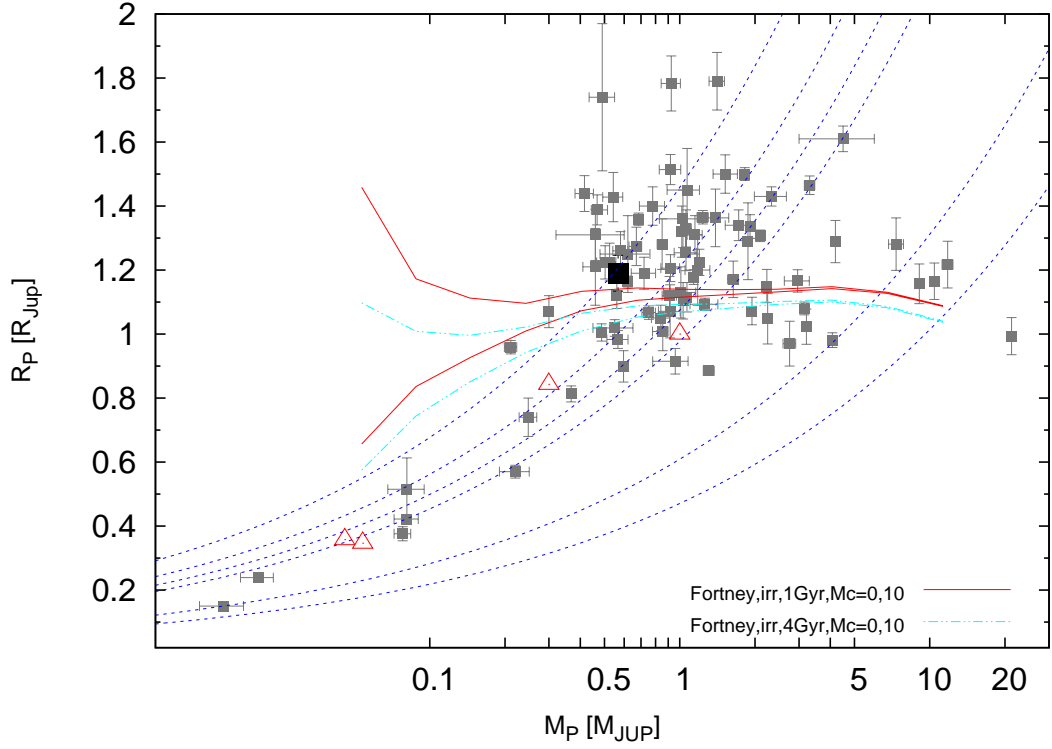}
\caption{
  Mass--radius diagram of known TEPs (small filled squares).
  \hatcurb\ is shown as a large filled square.  Overlaid are
  \citet{fortney:2007} planetary isochrones interpolated to the solar
  equivalent semi-major axis of \hatcurb{} for ages of 1.0\,Gyr
  (upper, solid lines) and 4\,Gyr (lower dashed-dotted lines) and core
  masses of 0 and 10\,\mearth\ (upper and lower lines respectively),
  as well as isodensity lines for 0.4, 0.7, 1.0, 1.33, 5.5 and
  11.9\,\gcmc\ (dashed lines).  Solar system planets are shown with
  open triangles.
  \label{fig:exomr}
}
\end{figure*}

\begin{figure*}[!ht]
\plotone{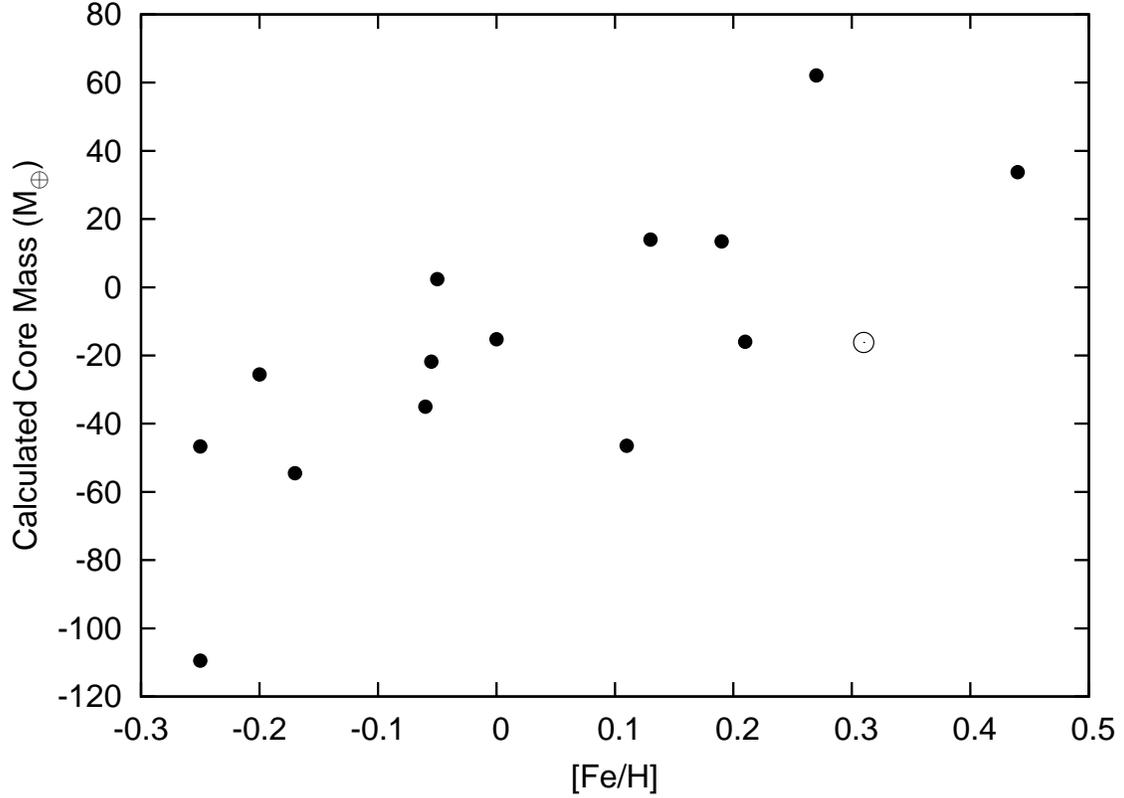}
\caption{
  Calculated planet core mass vs.~host star metallicity for planets
  with $0.4<M<0.6\,\mjup$.  The location of \hatcurb{} is shown by an
  open circle.  The core mass for each planet is determined by linear
  interpolation within the \citet{fortney:2007} planet model tables
  for the estimated age, mass, and solar-equivalent semimajor axis of
  the planet.  We adopt an age of 4.0\,Gyr or 0.3\,Gyr for systems
  with an estimated age greater or less than these limits.  Planets
  with a negative calculated core mass have radii that are too large
  to be accommodated by the models.  In this case the core mass is
  linearly extrapolated from the models, and provides a measure for
  the degree to which the observed radius disagrees with the models.
  \label{fig:core}
}
\end{figure*}

\reffig{exomr} compares \hatcurb{} to other known TEPs on a
mass-radius diagram.  Nothing unusual about it is immediately
apparent; its mass is between that of Saturn and Jupiter and its large
radius is typical of other short-period gas giant planets.  However,
this ``mediocrity'' gives us a rich sample against which to compare
it.  There are currently 19 other known transiting exoplanets with
$0.4<M<0.7\,\mjup$, which is a large enough sample to reveal trends in
their physical properties.  For example, \citet{enoch:2010} note the
apparent inverse relationship between planet radius and host star
metallicity, which is what would be expected if the heavy element
content of the planets (core masses) is proportional to host star
metallicity \citep{guillot:2006}.  \citet{hartman:2010} show a related
correlation, that between stellar metallicity and the planetary core
mass inferred from the \citet{fortney:2007} models.  \hatcurb{} is an
important addition to these analyses, as only XO-2b is more metal-rich
in this mass range.  \reffig{core} shows the location of \hatcurb{} in
the core mass vs.~\feh~relation.  We also note that \hatcurb{} shows
agreement with the previously illustrated correlations of planet
radius vs.~equilibrium temperature \citep[e.g.;][]{enoch:2010} and
orbital period vs.~planet mass (which was noted to be a marginally
statistically significant result by \citet{southworth:2010}).

\begin{figure*}[!ht]
\plotone{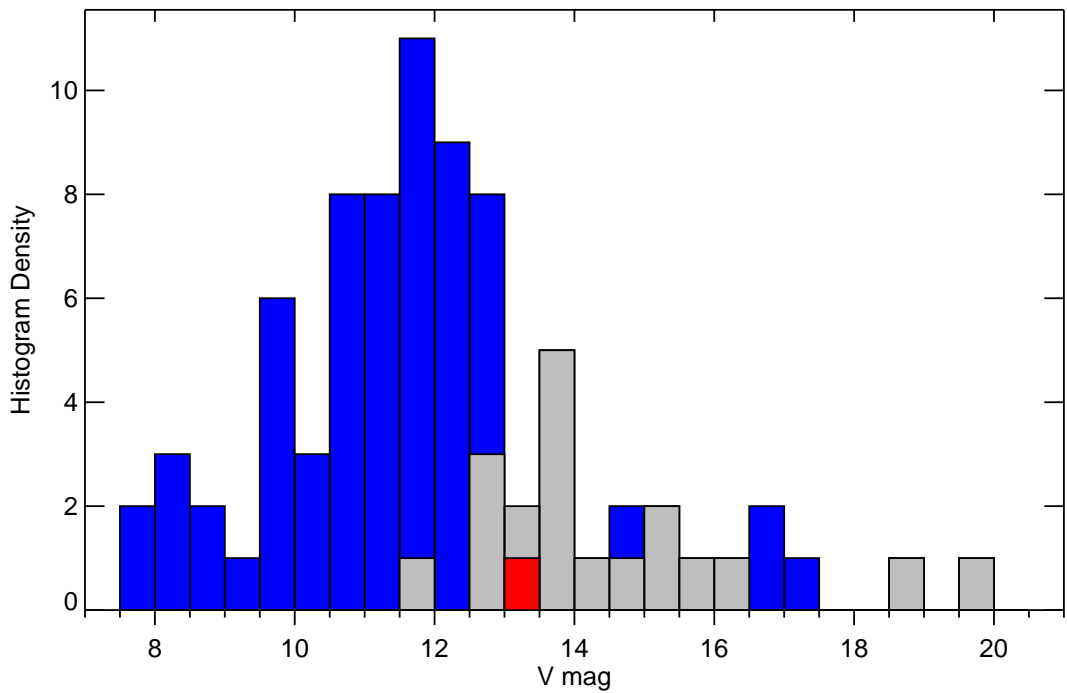}
\caption{
  Histogram of host star $V$ magnitudes. {\em Blue:} Host stars of
  transiting planets discovered from the ground.  {\em Gray:} Host
  stars of transiting planets discovered from space.  {\em Red:}
  \hatcur{}.
  \label{fig:vmag}
}
\end{figure*}

The characteristics of \hatcurb{} appear to be mostly typical of its
peers, although there is one aspect in which it is rather unusual for
a ground-based transit survey -- at $V=13.19$, it represents the
faintest system discovered by HATNet to date, and is one of only a
small handful of transiting planets discovered by ground-based surveys
orbiting stars fainter than $V=13$.  A histogram of host star
magnitudes is shown in \reffig{vmag}.  The fact that \hatcur{} is
faint is significant for a couple of reasons.  First, this
demonstrates that HATNet is capable of detecting planets orbiting
stars with $V>13.0$.  Second, and more importantly, follow-up of
fainter host stars will become more commonplace as we move forward.
It is obvious from the early results -- see \reffig{vmag} -- and the
known target lists that space-based surveys will provide a glut of
planets and planet candidates around fainter stars, and ground-based
surveys will eventually follow up their fainter candidates as well.
Systems such as \hatcurb{} afford the exoplanet community the
opportunity to hone our follow-up procedures for optimal performance
on faint stars.  For example, \hatcurb{} has already illustrated that
faint stars (and their low S/N) need to be treated carefully (if not
differently) when it comes to characterizing sky contamination.
Knowledge like this can help inform observing procedures and reduction
techniques in the future.

\clearpage

\acknowledgements
HATNet operations have been funded by NASA grants NNG04GN74G,
NNX08AF23G and SAO IR\&D grants.  Work of G.\'A.B.~and J.~Johnson were
supported by the Postdoctoral Fellowship of the NSF Astronomy and
Astrophysics Program (AST-0702843 and AST-0702821, respectively).  GT
acknowledges partial support from NASA grant NNX09AF59G.  We
acknowledge partial support also from the Kepler Mission under NASA
Cooperative Agreement NCC2-1390 (D.W.L., PI).  G.K.~thanks the
Hungarian Scientific Research Foundation (OTKA) for support through
grant K-81373.  This research has made use of Keck telescope time
granted through NOAO (A201Hr), NASA (N018Hr and N167Hr), and the NASA
Gemini-Keck time-exchange program (G329Hr).  This research has also
made use of the NASA/IPAC Infrared Science Archive, which is operated
by the Jet Propulsion Laboratory, California Institute of Technology,
under contract with the National Aeronautics and Space Administration.


\end{document}